\documentclass{aastex}
\usepackage{emulateapj5}
\usepackage{epsfig}

\begin{document}

\slugcomment{Submitted to the Astrophysical Journal}

\def\visuncal{{\tilde{V}}}
\def\vis{V}
\def\polexp{{Paper~IV}}
\def\paperfour{Paper~IV}
\def\polres{{Paper~V}}
\def\paperfive{Paper~V}
\def\paperone{{Paper~I}}
\def\papertwo{{Paper~II}}
\def\paperthree{Paper~III}
\def\check#1{{\bf #1}}
\def\stokes#1{S_#1}
\def\deg{^\circ}
\def\muK{~\mu{\rm K}}
\def\ngc{NGC~6334}
\def\rotm{{\rm RM}}
\def\snr{{s/n}}
\def\ra{{\rm R.A. }}
\def\dec{{\rm Dec.}}
\def\ran#1#2{{#1^{\rm h}}{#2^{\rm m}}}
\def\emode{$E$-mode}
\def\bmode{$B$-mode}
\def\rr{$\it RR$}
\def\ll{$\it LL$}
\def\rl{$\it RL$}
\def\lr{$\it LR$}
\def\te{$\it TE$}
\def\tb{$\it TB$}
\def\eb{$\it EB$}
\def\ee{$\it EE$}
\def\ete{$\it E/TE$}
\def\tte{$\it T/TE$}
\def\tetfive{$\it T/E/T5$}
\def\efivebfive{$\it E5/B5$}
\def\tete{$\it T/E/TE$}
\def\tetefive{$\it T/E/TE5$}
\def\tebtetbeb{$\it T/E/B/TE/TB/EB$}
\def\wmap{{\sl WMAP}}
\def\copolarFrac{$53\%$}
\def\cxpolarFrac{$61\%$}
\def\austral{Austral}
\def\insnum{{\bf NaN}}
\def\threeyear{3-year}

\title{DASI Three-Year Cosmic Microwave Background Polarization Results}

\author{E.\ M.\ Leitch$^1$, J.\ M.\ Kovac$^2$, N.\ W.\ Halverson$^3$, 
J.\ E.\ Carlstrom$^{1,4,5}$, C.\ Pryke$^{1,5}$ \\ and M.\ W.\ E.\ Smith$^1$}

\affil{$^1$Kavli Institute for Cosmological Physics,
Department of Astronomy \& Astrophysics, 
University of Chicago,
5640 South Ellis Avenue, 
Chicago, IL 60637}

\affil{$^2$California Institute of Technology,
1200 East California Blvd,
Pasadena, CA 91125}

\affil{$^3$University of California, Department of Physics \& Space Sciences Laboratory, Berkeley, CA 94720}

\affil{$^4$Department of Physics, 
University of Chicago,
Chicago, IL 60637}

\affil{$^5$Enrico Fermi Institute, 
University of Chicago,
Chicago, IL 60637}

\begin{abstract}
We present the analysis of the complete \threeyear\
data set obtained with the
Degree Angular Scale Interferometer
(DASI) polarization experiment, operating from the Amundsen-Scott
South Pole research station.
Additional data obtained at the end of the 2002 \austral\ winter
and throughout the 2003 season were added to the data from
which the first detection of polarization of the cosmic microwave
background radiation was reported. The analysis
of the combined data supports, with increased statistical power,
all of the conclusions drawn from the initial data set.
In particular, the detection of \emode\
polarization is increased to $6.3\sigma$ confidence level,
\te~cross-polarization is detected at $2.9\sigma$, and
\bmode~polarization is consistent with zero, with an upper
limit well below the level of the detected \emode\ polarization. The 
results are in excellent agreement
with the predictions  
of the cosmological model that has emerged
from CMB temperature measurements. 
The analysis also demonstrates that contamination of the data by
known sources of foreground emission is insignificant.

\end{abstract}


\section{Introduction}

The use of the cosmic microwave background (CMB) as a 
precision probe of the composition and dynamics of the universe
presumes the correctness of the theoretical framework by which we
understand its origin and evolution.
With the detections of CMB polarization anisotropy by DASI \citep[hereafter
\polres]{kovac02} and \wmap\ \citep{kogut03a} that framework itself 
has passed a
critical test,
as the
presence of polarization is a generic requirement of the physics of
recombination \citep[for a recent review see][]{hudodelson02}.  The intrinsic polarization of the CMB reflects the
local radiation field anisotropy, specifically the quadrupole moment, at the surface of last
scattering, 14 billion years ago.  Within the context of
the standard cosmological model, in which peaks in the CMB angular
power spectrum are interpreted as the signature of acoustic
oscillations seeded by primordial density fluctuations, current
temperature measurements lead to specific predictions for the shape 
and amplitude of
the polarization power spectrum and the
temperature-polarization cross power spectrum.  Furthermore, it is a firm
prediction that density fluctuations should create only \emode, i.e.,
curl-free, polarization patterns on the sky.

In \polres, we reported the high signal to noise (\snr) detection of
\emode\ polarization in the CMB with DASI.  \citet{kogut03a}
reported 
the detection of the \te~correlation on larger angular
scales by \wmap.
The DASI and \wmap\ measurements
underscore the power of combining sub-orbital
experiments, with their ability to achieve exquisite sensitivity
over
small areas of sky, with space experiments, whose sky
coverage allows them to approach fundamental sample variance limits.
These measurements provide fuel for a host of new experiments aimed at
probing the detailed polarization spectrum of the CMB, from the
largest scales, which may bear the signature of inflationary gravitational
waves, to the smallest, where the CMB is expected to record the
history of structure formation.

In \polres, we presented the analysis of the initial DASI polarization
data set obtained during 2001 and the first part of the 2002 season;
modifications to the instrument to enable polarization measurements,
as well as the observations and calibration of the data, were
described in \citet[hereafter Paper IV]{leitch02b}. In this paper we
present the analysis of the \threeyear\ DASI polarization data set,
which comprises the initial data (271 days of observations) and the
additional data collected during the end of the 2002 season and over
the entire 2003 season (191 days in all).
The end of the 2003 season marked the
conclusion of interferometric observations with DASI.

\section{Observations}

\subsection{Instrument Review}

The use of interferometers, and of DASI in particular, to measure the 
angular power
spectrum of CMB anisotropy is described extensively
in \citet[hereafter \paperone]{leitch02a} and in \paperfour.
In short, an interferometer detects correlations between pairs of
antennas, i.e., {\it baselines}, yielding measurements which are
largely insensitive to uncorrelated noise, and which map simply into
Fourier components of the sky brightness distribution, longer
baselines measuring higher spatial frequencies.  As they operate
natively in the Fourier plane, interferometers are ideally suited to
measurement of the power spectrum.

DASI is a 13-element, co-planar interferometric array operating in
ten 1-GHz bands spanning $26-36~$GHz.
The telescope sits atop an eleven meter
tower located at the Amundsen-Scott South Pole research station.
DASI was initially configured
for measurements of
the CMB temperature anisotropy on angular scales corresponding to multipoles $140 < l < 900$ (\paperone). 
These observations were completed during the 
2000 \austral\ winter and the results reported in 
\citet[hereafter \papertwo]{halverson02} and \citet[hereafter \paperthree]{pryke02}.
As described in \paperfour, several changes were made for
the polarization observations, including the installation
of a 
large
reflective radiation shield surrounding the telescope. The shield
geometry ensures that all lines of sight from the telescope in the direction of the station buildings, the
horizon and ground, are reflected to the sky.
To provide polarization sensitivity, 
the thirteen DASI receivers 
were each equipped with cooled, broadband achromatic polarizers which
can be mechanically set to accept either left or right-circularly polarized light.  Within
each 1-hour interval of observation, the polarizers were stepped between
right ($R$) and left ($L$) positions to acquire data in all four Stokes states,
which we refer to as {\it co-polar} (\rr, \ll) and {\it
cross-polar} (\rl, \lr) data.

A sun shield, designed to extend the polar observing season to include
periods just before sunset and after sunrise, was installed for the
2002 season.  The shield was found to contaminate the shortest
baselines by secondary reflection at times when the moon was above the
horizon, and was removed for the final season.  (Note however that the
large ground shield remained in place during all three seasons.)
Apart from the installation and removal of the sun shield, and replacement of
miscellaneous hardware during routine maintenance of the telescope,
the instrument was identical throughout the three seasons discussed
here.

\subsection{Observing Strategy}

\begin{figure*}[t]
\epsscale{1.3}
\plotone{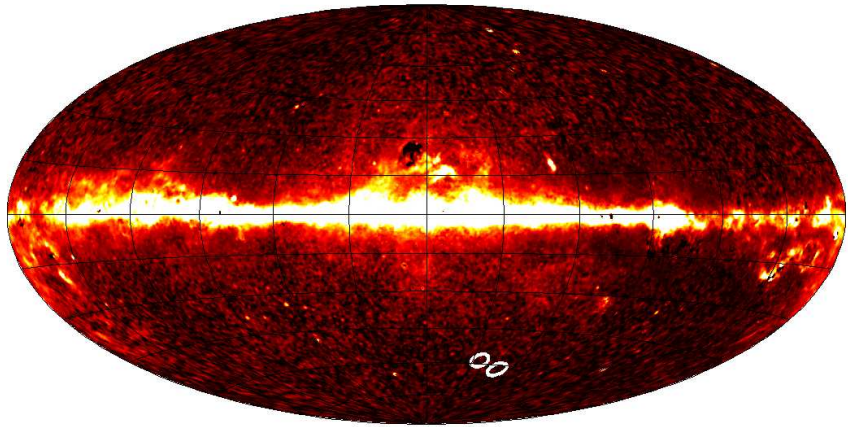}
\caption{Polarization field locations.  DASI fields are marked by
white circles, plotted over the Ka-band synchrotron map derived from
the \wmap\ first-year data \citep{bennett03b}.  The color scale is
linear and has been clipped at one thirtieth of the maximum to allow
the structure at high Galactic latitudes to be seen.}
\label{fig:wmap}
\end{figure*}

The polarization experiment consisted of deep observations of two
$3\fdg4$ FWHM fields, at Right Ascension 23$^h$30$^m$ and 00$^h$30$^m$, and
Declination $-55\deg$ (J2000) (see Figure~\ref{fig:wmap}). Details of the
field selection are given in \polexp.  The fields were observed
continuously at constant elevation (one of the unique advantages of
observing from the South Pole), interspersed with daily observations of a
calibrator source.  A variety of cross-checks were built into the
experiment design. Observations of the two fields were interleaved on
1-hour timescales to ensure that each was observed over the same
azimuth range, allowing effective removal of any residual ground
signals by differencing the data.  The configuration of
the array elements is threefold symmetric and the entire array is fixed to a
faceplate which can be rotated along the line of sight. On 24-hour
timescales, the observations alternated between two different orientations
of the faceplate, offset by 60$\deg$, in which the same Fourier
components are measured, but by different
pairs of antennas. 

Throughout the three seasons of observations, the observing strategy,
as described in \paperfour, remained identical.

\vspace{0.4in}
\subsection{Data Cuts}

A variety of cuts, discussed extensively in \polexp, are applied to
the data before input to the likelihood analyses; cut criteria and
thresholds for the combined three-season data were the same as for the
data presented in \polexp~and \polres.  During the 2002 season, when
it was discovered that the short-baseline data were affected by
reflections of the moon off the sun shield, a more stringent cut on
the elevation of the moon was applied than during the 2001 season.
During the 2003 season, for which the sun shield had been removed, cuts on
the moon were relaxed to the 2001 levels.
Early in
the 2003 season, the polarizer assembly for one of the DASI receivers began to slip, and data associated with this receiver were 
therefore excised for the
entire season, removing $\sim15\%$ of the 2003 data.

Of the three-season data 
taken with the sun 
below the horizon, approximately \copolarFrac~of the co-polar
data, and \cxpolarFrac~of the cross-polar data passed the data cuts.
Cut fractions were dominated by cuts on the elevation of the moon, and
by cuts on the complex gains of multipliers in the correlator (see
\polexp).

\subsection{Calibration}

\label{sec:calibration}

The instrumental response of each baseline of an interferometer is the
complex multiplication of the individual antenna responses.
 As a result, the
polarization calibration, up to an absolute phase offset, can be
derived from observations of an unpolarized source (see \polexp).  The remaining
phase offset was determined by observing an unpolarized source through
linearly polarizing wire grids attached to the front of each receiver.  
These offsets had
previously been measured in 2001 August and 2002 February.  The
wire-grid calibrations were performed again in 2003 March and 2003 October,
and results from all four epochs agree to within
the measurement uncertainties.

Leakage from one polarization state to the other will mix CMB power
from $T$ into \emode s and \bmode s, and must be included in
the analysis.  Ideally these should be kept as small as possible, but
even significant leakages can be accounted for if they are stable.  The
leakages can be separated into a component due to the polarizers
(referred to as {\it on-axis} leakages in \polexp), and a
beam-dependent component which derives from the inherent properties of
the optics (referred to as {\it off-axis} leakages in \polexp).  The
latter were characterized in \polexp, and are not expected to change.
The on-axis leakages for DASI were determined from observations of a
bright unpolarized source in 2001 August, 2002 April, 2002 July and
2003 August.  On-axis leakages of total intensity into the
cross-polarized data were found to be less than
$1\%$ at all but the highest frequency bands (see \paperfour) and were
consistent from year to year.

\section{Results}

\subsection{Noise Model and Consistency Tests}
\label{sec:consistency}

For both the $\chi^2$ consistency tests and likelihood analyses that
follow, the calibrated, field-differenced, and co-added visibility
data are arranged in a data vector that is considered to be the sum of
signal and noise components, $\mathbf{\Delta} = \mathbf{s} +
\mathbf{n}$. An accurate characterization of the instrument noise in
the data vector is critical for estimating the faint
polarization signal. We calculate various statistics from the
raw data to test our assumptions about the noise model: namely, that
the noise is Gaussian, that the noise variance is well approximated by
the variance in the 8.4-s samples (the fundamental data rate) over
1-hr intervals before field differencing, and that the noise is
uncorrelated between data vector elements.

Details of the noise model tests and results for the initial data set
are given in \polres.  We find similar results for the \threeyear\
data. Three different types of short-timescale (1-hr) variance
estimators are in excellent agreement with one another, and also with
the variance derived from the scatter of 1-hr binned visibilities over
the \threeyear\ span of the data set. In \polres, we scaled the noise
by a factor of 1\% (twice the maximum discrepancy among the various
noise estimators described above) to test the robustness of our noise
estimate and found no significant change in the likelihood results.

We construct a data covariance matrix for each day of observation to
test the assumption that the noise covariance matrix is diagonal, and
to form a cut statistic based on correlated atmospheric noise. Data
covariance matrices for all good-weather days are averaged, to test
for noise correlations persistent throughout the data set. We find
significant correlations in 0.017\% of the off-diagonal terms. These
are similar in number and magnitude to those previously described in
\polres, where we found that adding the measured correlations to the
noise model had a negligible effect on the final results.

We construct a $\chi^2$ statistic on sum and difference data vectors
for various splits and subsets of the data in which the same signal is
expected, as a simple but powerful test of the noise model, data
consistency, and presence of signal in the data. The statistic is
given by
$\chi^{2}=\mathbf{\Delta}^{t}\mathbf{C}_{N}^{-1}\mathbf{\Delta}$,
where $\mathbf{\Delta} = \left( \mathbf{\Delta_1} \pm
\mathbf{\Delta_2} \right) / 2$ is the sum or difference data vector,
and $\mathbf{C}_{N}= \left( \mathbf{C}_{N1}+\mathbf{C}_{N2} \right)
/4$ is the corresponding noise covariance matrix.  We assess the
significance of a $\chi^2$ value by calculating the probability to
exceed (PTE) that value in the cumulative distribution function.

Following the procedure in \polres, we split the data by epoch,
azimuth range, faceplate position, and Stokes state (for co-polar
observations) for each of the three years of data individually, for
each year versus the other two, and for the full \threeyear\
dataset. For each split, we also examine various subsets of the data
vector, including co-polar and cross-polar data, data in distinct
multipole ranges and the high expected \snr\ eigenmodes. In
all, over 400 $\chi^2$ tests were performed on various splits/subsets
of the data.

The distribution of $\chi^2$ values for the differenced polarization
data shows no evidence for residual signal in excess of the noise,
giving a uniform distribution of PTE values with no apparent trends or
outliers.  In the 2003 temperature data, several subsets which isolate
high expected \snr\ eigenmodes formed from the shortest baselines
compared at different faceplate rotations yield high $\chi^2$ values,
the most significant of which gives a PTE = 0.0004.  Because this
discrepancy appears in an isolated subset of data which contains an
extremely high signal, it is suggestive of a small shift in
instrumental response.  Such a shift at the level suggested by these
tests would not significantly affect the data.

For the sum data vectors, a signal is detected with high significance
in all of the co-polar splits and subsets, as well as in many of the
high \snr\ subsets of the cross-polar data. In addition to the split
data tests, we construct $\chi^2$ statistics from the highest \snr\
eigenmodes of the unsplit data vectors, as a sensitive check for the
presence of signal, as discussed in \polres.
In the cross-polar data, the number of expected \snr\ $> 1$
polarization eigenmodes has increased from 34 in the initial data set
to 60 for the \threeyear\ data set (taking into account off-axis
leakage terms in the noise covariance matrix $\mathbf{C}_{N}$, see
\polres).  For these eigenmodes, we find $\chi^2 = 141.3$, which
corresponds to a PTE of $1.66\times 10^{-8}$ --- a high-significance
detection of a polarized signal.

The results of both the $\chi^2$ consistency tests and the noise
model tests demonstrate that the noise is well characterized and
Gaussian, and that the data calibration is consistent over the three years
of observations. The sum data vector $\chi^2$ tests on both the
co-polar and cross-polar data indicate the detection of signal with
high-significance, using a simple test that is independent of the more
sophisticated likelihood analysis results that follow.

\subsection{Systematic uncertainties}
\label{sec:systematics}

In \polres, we investigated the impact of possible sources of
systematic uncertainties on the results of the likelihood analysis. In
addition to the uncertainties in the noise model discussed above, we
considered systematic uncertainties in the absolute cross-polar phase
offsets, in the on-axis and off-axis leakages (see
\S\ref{sec:calibration} for definitions), and in the pointing of the
telescope.  In each case, re-analyzing the data with the
systematic errors set to the limits of their allowed ranges led to
insignificant shifts in the results.

The overall uncertainty of the DASI absolute calibration, derived from
measurements of external thermal loads, transferred to an astronomical
calibrator, was estimated in \papertwo\ to be 8\% (1$\sigma$),
expressed as a fractional uncertainty on the $C_l$ bandpower (4\%
in $\Delta T/T$), and applies equally to the temperature and
polarization data.

\subsection{Likelihood Results}
\label{sec:lhresults}

The same likelihood analysis formalism described in \polres\ was
used to
analyze the DASI \threeyear\ polarization data set. We place constraints on
a given set of parameters by characterizing the shape of the
likelihood function, i.e., the probability of our data vector $\mathbf{\Delta}$, given a
set of parameters $\mathbf{\kappa}$ which describe the signal,

\begin{eqnarray}
\label{eq:like}
\nonumber
L\left(  \mathbf{\kappa}\right) &=& P\left( \mathbf{\Delta} | \mathbf{\kappa} \right)  \\
&\propto& \det\left(  \mathbf{C}\left(
\mathbf{\kappa}\right)  \right)  ^{-1/2}\exp\left(  -\frac{1}{2}%
\mathbf{\Delta}^{t}\mathbf{C}\left(  \mathbf{\kappa}\right)  ^{-1}%
\mathbf{\Delta}\right),
\end{eqnarray}
where 
$\mathbf{C}\left(\mathbf{\kappa}\right)$ is a model for the covariance matrix of
the data vector.
This matrix contains the expected 
contributions of the instrument noise and
the cosmological signal, as seen 
through the particular filter of the
experiment. Details of the construction of
$\mathbf{C}\left(\mathbf{\kappa}\right)$ for the DASI polarization
data set are given in \polres\ and in \citet{kovac_thesis}.

We report the maximum likelihood (ML) values of the parameters
$\mathbf{\kappa}$, found using an iterated quadratic estimator
technique \citep{bond97}. The shape of the likelihood function in the
vicinity of the ML peak is characterized using direct grid evaluation
for those analyses with four or fewer parameters, and using a Markov
chain algorithm for higher dimensional analyses \citep{christensen01}.
Confidence intervals for individual parameters are obtained by
marginalizing (integrating) the likelihood function over the remaining
parameters, and reporting the equal-likelihood bounds which enclose
68\% of the marginal likelihood distribution (the so-called highest
posterior density (HPD) interval). These are given in the text in the
format: ML (HPD-low to HPD-high). For parameters which are
intrinsically positive, we consider placing (physical) upper limits by
marginalizing the likelihood distribution as before, but excluding the
unphysical negative values.  We then test if the 95\% integral point
has a likelihood smaller than that at zero; if it does, we report an
upper limit rather than a confidence interval.
We also include in tabulated results
the uncertainty estimates and parameter correlation coefficients
obtained by evaluating the Fisher curvature matrix $F$ at the peak of
the likelihood function.

Our goodness-of-fit statistic, as described in \polres, is the
logarithmic ratio of the maximum of the likelihood to its value for
some hypothetical model ${\cal H}_0$ described by parameters $\kappa_{0}$,
\[
\Lambda({\cal H}_0)\equiv-\log\left(  \frac{L\left(  \kappa_{ML}\right)  }{L\left(
\kappa_{0}\right)  }\right)\,.
\]
In the limit where the $\kappa_{ML}$ parameter uncertainties are
normally distributed, the statistic reduces to
$\Lambda=\Delta\chi^2/2$. Where possible, we have tested the
distribution of $\Lambda$ using Monte Carlo simulations and found
$\Delta\chi^2/2$ to be an excellent approximation. We therefore use
the well-known $\chi^2$ cumulative distribution function to compute
the probability that, assuming a given hypothetical model ${\cal H}_0$
to be true, we would obtain a value for $\Lambda({\cal H}_0)$ that
exceeds the observed value.  This is quoted in the text and tabulated
results as a PTE and, where appropriate, as a significance of
detection, in units of $\sigma$.

For the \threeyear\ data set, we perform all of the analyses reported
in \polres\ for the initial data set. In addition, we perform two new
analyses to constrain the level of contamination from synchrotron or
radio point source emission.  The separate likelihood analyses
constrain different aspects of the signal in the polarization data, in
the temperature data, or in both analyzed together. The numerical
results are given in Tables \ref{tab:cxpolar_like},
\ref{tab:copolar_like} and \ref{tab:joint_like}, respectively.
Closely following the format from \polres, we now briefly discuss each
analysis in turn.

\subsubsection{$E/B$ Analysis}
\label{sec:EB}

In this analysis, two single-bandpower parameters are used to
characterize the amplitudes of the $E$ and $B$ polarization spectra.
DASI has an instrumental response to
$E$ and $B$ that is symmetric and nearly independent.  Although the B
spectrum is not expected to have the same shape as the $E$ spectrum,
we choose the same shape for both spectra to preserve this symmetry in
the analysis.

Following \polres, we consider three {\it a priori} shapes to
determine which is most favored by the data: the
concordance\footnote{A $\Lambda$CDM model with flat spatial curvature,
5\% baryonic matter, 35\% dark matter, 60\% dark energy, and a Hubble
constant of 65~km~s$^{-1}$~Mpc$^{-1}$, $(\Omega_b = 0.05, \Omega_{cdm}
= 0.35, \Omega_\Lambda = 0.60, h = 0.65)$ and the normalization
$C_{10} = 700 \mu K^2$. Using the WMAPext parameters 
changes the results of our $T$, $E$ and $TE$ amplitudes
by less than 3\%.}
 $E$ spectrum shape, and two alternatives:
a flat $l(l+1)C_l \propto \rm{constant}$ spectrum, and a power law
$l(l+1)C_l \propto l^2$ spectrum.  For each of these three cases, the
point at $E=0,\ B=0$ corresponds to the same zero-polarization {\it
nopol} model, so that the likelihood ratios $\Lambda({\it nopol})$ may
be compared directly to assess the relative likelihoods of the
best-fit models.  For the flat case, the ML flat bandpower values are
$E=6.2 \mu K^2, B=0.0 \mu K^2$, with $\Lambda({\it nopol}) = 10.58$.
For the $l^2$ power law case, the ML values are $E=4.2 \mu K^2, B=0.0
\mu K^2$ ($l(l+1)C_l/2\pi$ at $l=300$), with $\Lambda({\it nopol}) =
15.41$.  For the concordance shape, the ML values are $E=0.73,B=0.03$
in units of the concordance $E$ spectrum amplitude, with $\Lambda({\it
nopol}) = 20.10$.  Although each of these analyses shows strong
evidence for an $E$ polarization signal, the likelihood of the best
fit model in the concordance case is a factor of 110 and 14,000 higher
than those of the $l^2$ power law and flat cases, respectively.
The data clearly prefer the concordance
shape, which we therefore use for the $E/B$ and other single bandpower analyses.

\begin{figure*}[t]
\begin{center}
\epsfig{file=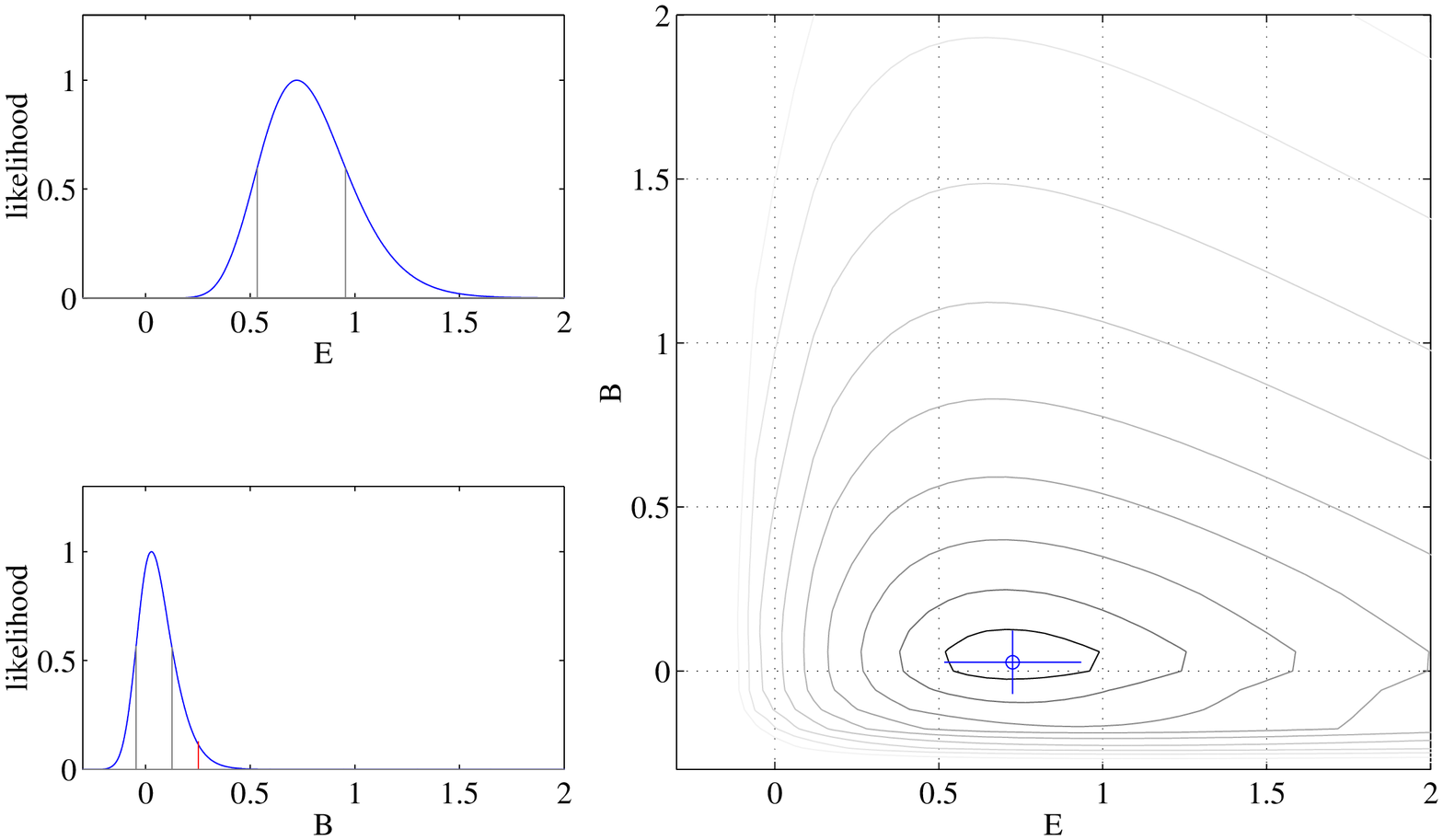,width=\textwidth}
\end{center}
\caption{Results from the two-parameter shaped bandpower $E/B$ polarization analysis.
We assume the $E$-mode power spectrum shape predicted for
the concordance model, and the units of amplitude
are relative to that model.
The same shape is assumed for the $B$-mode spectrum.
(right panel) The point shows the maximum likelihood value with
the cross indicating Fisher matrix errors.  Likelihood contours are placed at levels
$\exp(-n^2/2)$, $n=1,2...$, relative to the maximum, i.e., for a normal distribution,
the extrema of these contours along either dimension would give the marginalized
$n$-sigma interval.
(left panels) The corresponding single-parameter likelihood
distributions marginalized over the other parameter.
Note the steep fall in likelihood toward low-power values; this likelihood shape (similar to a $\chi^2$ distribution)
is typical for positive-definite parameters for which a limited number of high signal/noise modes
are relevant.
The grey lines enclose 68\% of the total likelihood.
The red line indicates the 95\% confidence upper limit on $B$-mode power.
}
\label{fig:EBplot}
\end{figure*}

Figure~\ref{fig:EBplot} illustrates the result of the $E/B$ concordance-shape polarization analysis.
The maximum likelihood value of $E$ is 0.73 (0.53 to 0.95 at 68\% confidence).
For $B$, the result should clearly be regarded as an upper limit;
95\% of the $B > 0$ likelihood (marginalized over $E$) lies below 0.25.

The data are highly incompatible with the no polarization hypothesis.
The likelihood ratio $\Lambda({\it nopol}) = 20.10$
implies a probability that the data are consistent with the
zero polarization hypothesis of $\rm{PTE} = 1.86 \times 10^{-9}$, assuming that the uncertainties in $E$ and $B$ are normally distributed, as discussed above.
Marginalizing over $B$, we find $\Lambda(E=0) = 19.65$, corresponding 
to detection of $E$-mode polarization at a PTE of $3.64 \times 10^{-10}$,
or a significance of $6.27 \sigma$.

The likelihood ratio for the concordance model 
gives $\Lambda(E=1,\,B=0)=0.68$, for which the PTE is 0.51.
We conclude that the data are consistent with the concordance model.

\subsubsection{\efivebfive}
\label{sec:E5B5}

\begin{figure*}[t]
\begin{center}
\epsfig{file=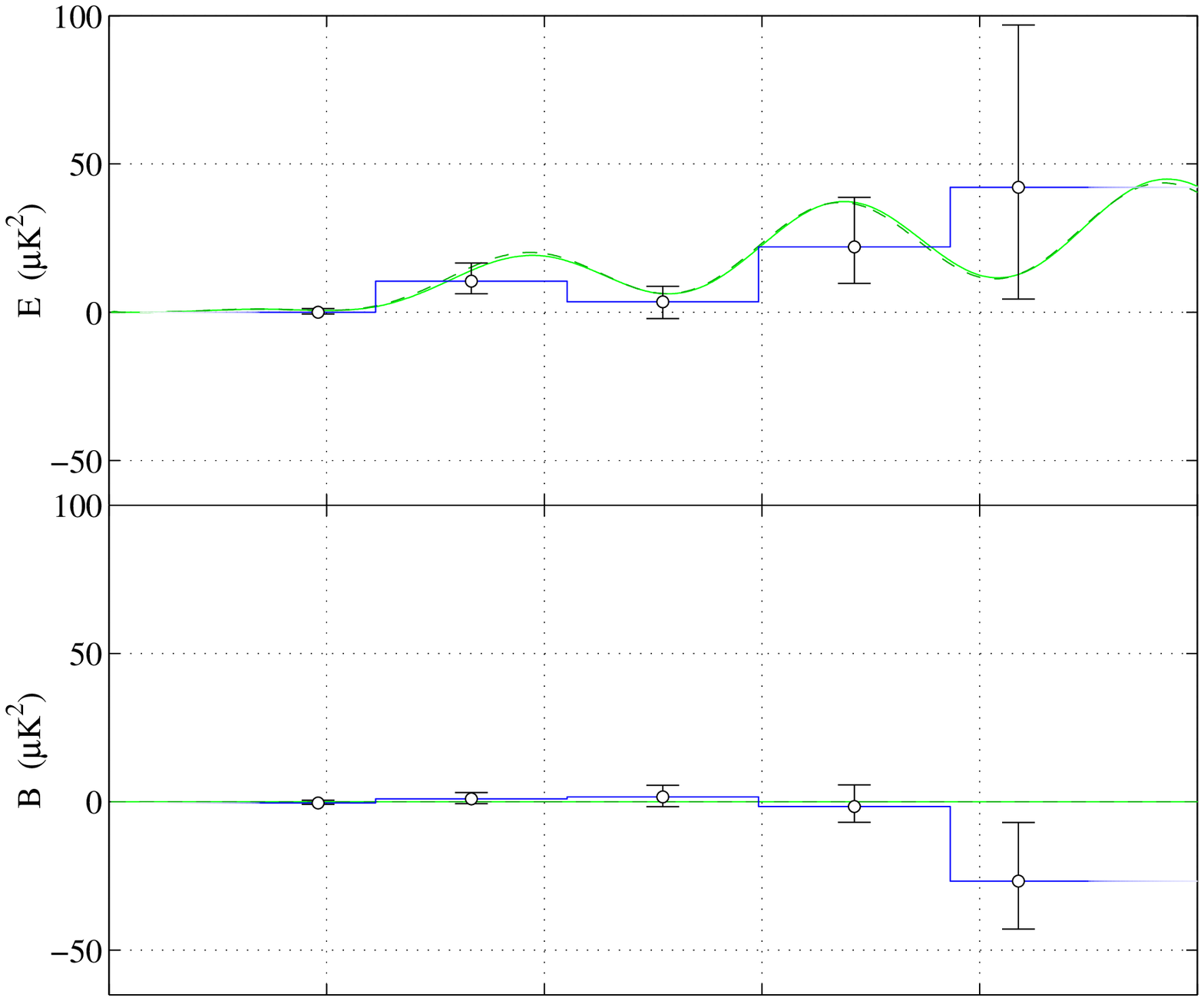,width=4.5in}
\vfill
\epsfig{file=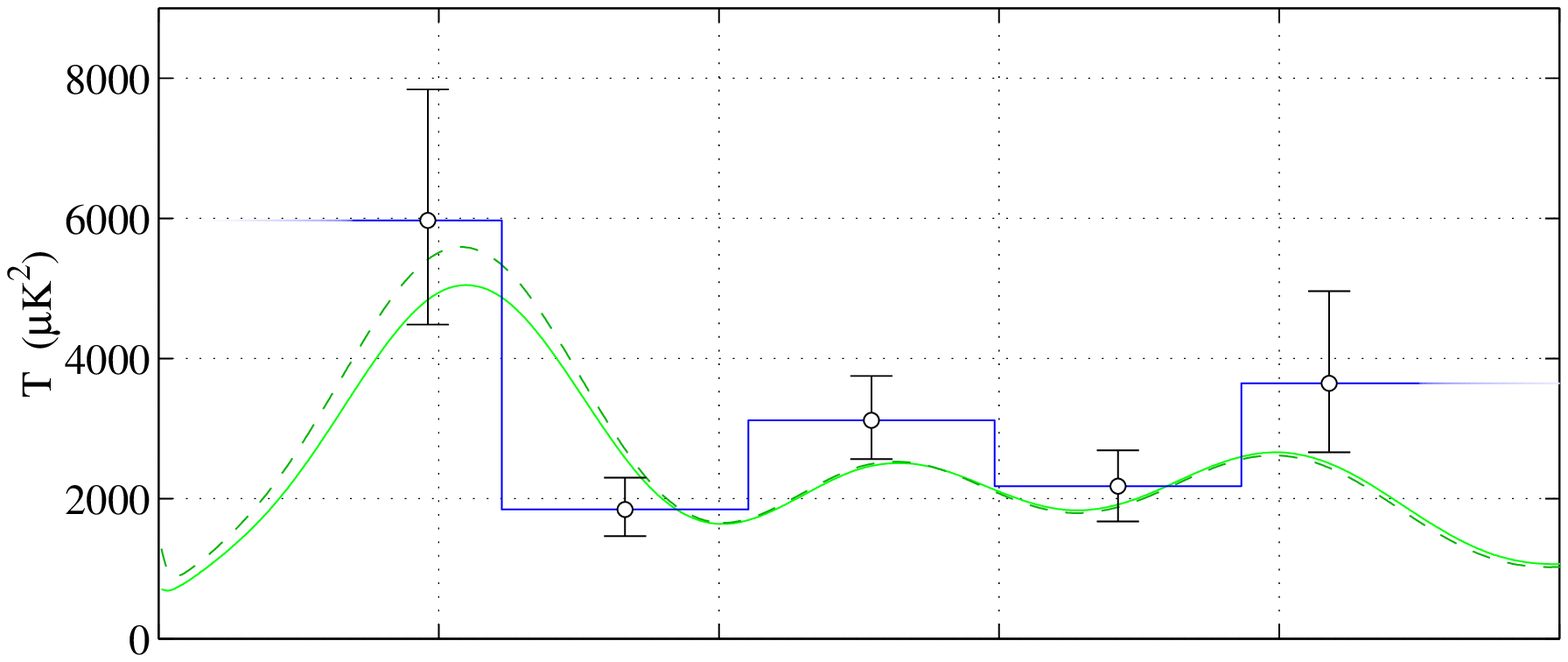,width=4.5in}
\vfill
\epsfig{file=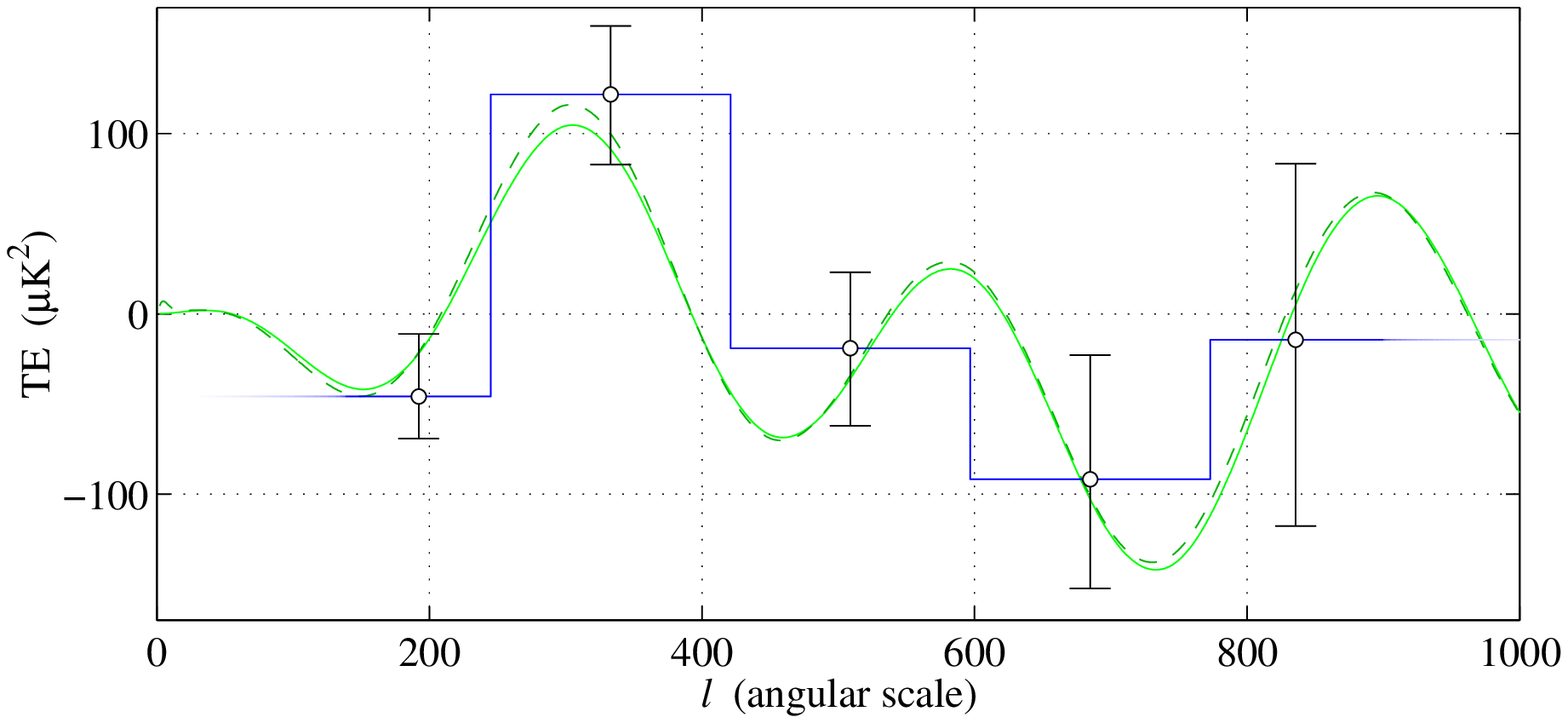,width=4.5in}
\end{center}
\caption{Results from several likelihood analyses. 
The top two panels show the ten-parameter $E5$/$B5$ polarization
analysis, the third panel shows the $T5$ temperature analysis and the
bottom panel shows the five \te\ bands from the \tetefive\ joint
analysis.  All results are presented in flat bandpower units of
$l(l+1)C_{l}/{2\pi}$.  The blue lines show the piecewise-continuous,
flat-bandpower model for the maximum likelihood parameter values, with
error bars indicating the 68\% central region of the likelihood of
each parameter, marginalizing over the other parameter values,
analogous to the grey lines in the left panels of
Figure~\ref{fig:EBplot}.  In each case, the solid green line is the
concordance model used in \polres\ and this paper, and the dashed
dark-green line is a best-fit model for the \wmap ext data from
\citet{spergel03}.}
\label{fig:fivebandTEBTE}
\end{figure*}

This analysis parameterizes the $E$ and $B$-mode spectra using five,
piecewise-continuous, flat bandpowers for each.  Results of this
ten-parameter analysis are shown in Figure~\ref{fig:fivebandTEBTE}.  The
$l$-ranges of the five bands, provided in Table
\ref{tab:cxpolar_like}, are unchanged from the equivalent analysis
presented in Paper V, with corresponding window functions similar to
those previously reported.

To test for consistency with the concordance model, we calculate the
expectation value for that model for each of the five bands, yielding
$E$=(0.8,14,13,37,16) and $B$=(0,0,0,0,0) $\muK^2$.  The ratio of the
likelihood at this point in the ten dimensional parameter space to the
maximum likelihood gives $\Lambda=3.21$, which for ten degrees of
freedom results in a PTE of 0.78, indicating that our data are
consistent with the expected polarization parameterized in this
way. The \efivebfive\ results are also highly inconsistent with the
zero polarization {\it nopol} hypothesis, with $\Lambda=19.47$, which
for 10 parameters yields ${\rm PTE} = 2.6 \times 10^{-5}$.  This
likelihood ratio is close to the value $\Lambda({\it nopol}) = 20.10$
obtained for the best-fit $E/B$ concordance-shaped model in the
previous section, indicating that the two best-fit models, though
obtained with different numbers of parameters, offer similarly likely
descriptions of the polarization signal.

\subsubsection{$E$/$\beta_E$}

This analysis parameterizes the $E$-mode polarization signal as above, as well as the
frequency spectral index $\beta_E$ of this signal relative to the CMB
(such that $\beta = 0$ corresponds to a 2.73~K Planck spectrum).  As
expected, the results for the $E$-mode amplitude are very similar to
those for the $E/B$ analysis described above.  The result
for the spectral index is $\beta_E = 1.39$ ($-0.21$ to 2.92). The
index is consistent with CMB, as $\Lambda(\beta_E=0)=0.372$ for a
PTE = 0.39. It is inconsistent with synchrotron emission, which is rejected
at $2.65\sigma$ ($\Lambda(\beta=-2.9)=3.52$ for a PTE = 0.008).
Although the constraint is not strong, 
it is nevertheless
interesting in the context of assessing the level of possible
contamination by non-thermal emission, as discussed in
\S\ref{sec:discussion}.

\subsubsection{$E/sync$}
\label{sec:Esync}

\begin{figure*}[t]
\begin{center}
\epsfig{file=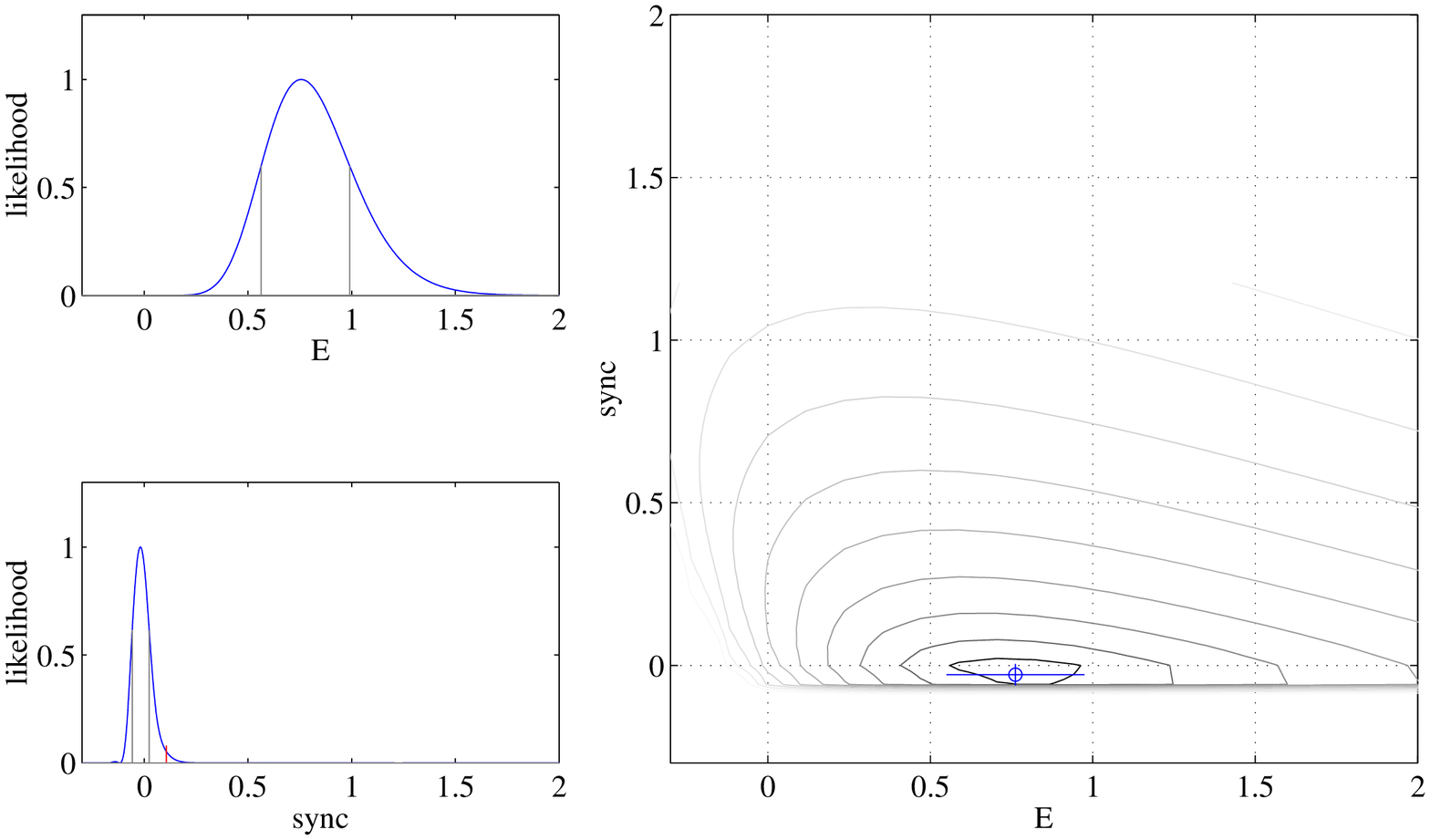,width=0.8\textwidth}
\epsfig{file=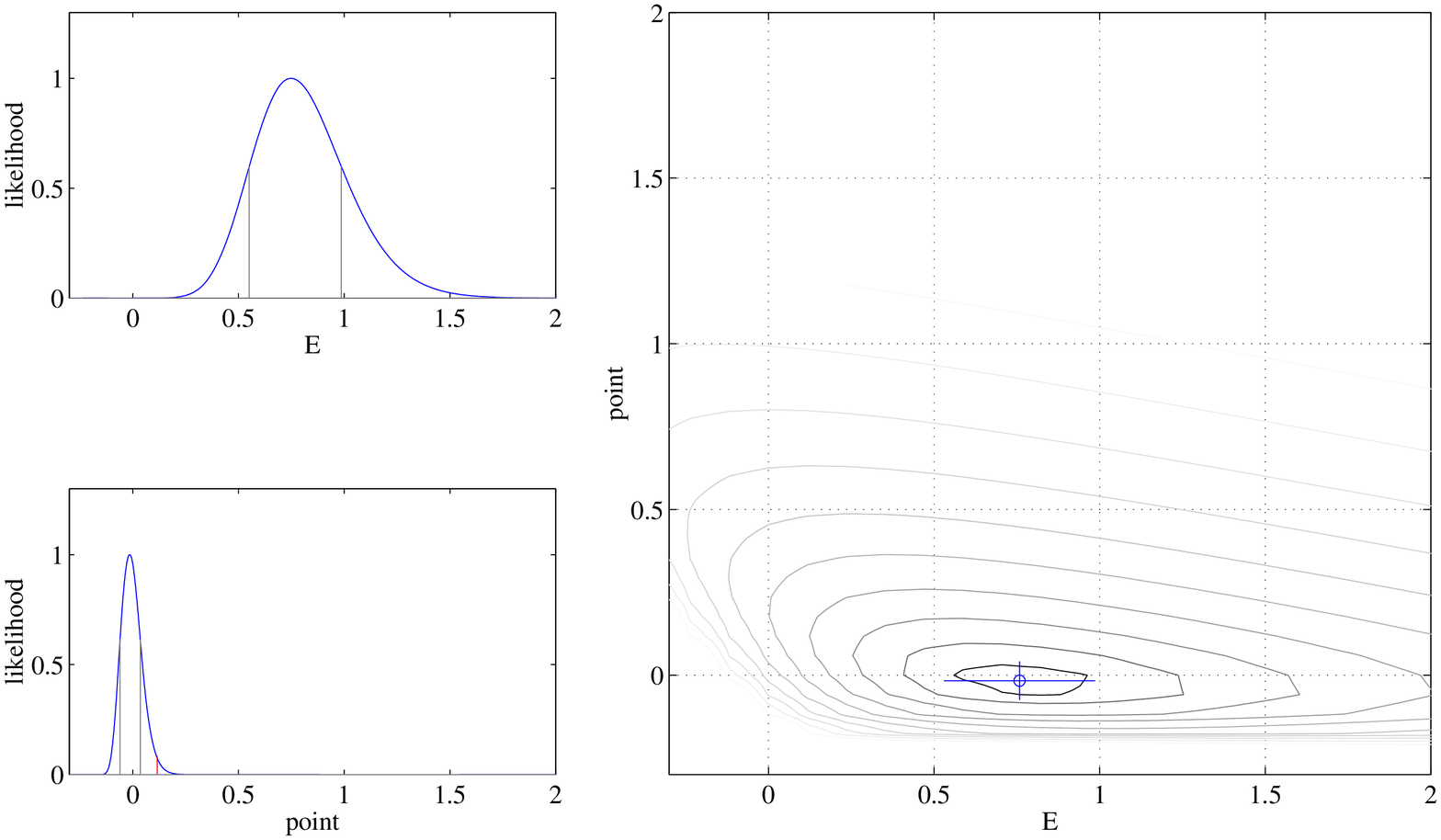,width=0.8\textwidth}
\end{center}
\caption{Results of $E/sync$ and $E/point$ two-component polarization
analyses. 
These two analyses, performed to constrain the level of foreground emission
in the data set, both use for their first component the same CMB $E$ power spectrum
as in the $E/B$ concordance-shape polarization analysis (Figure~\ref{fig:EBplot}).
For the $E/sync$ analysis shown in the upper three panels, the second component is
a model for polarized synchrotron emission with $E = B$, flat power spectra and
temperature spectral index $\beta = -2.9$ (see \S\ref{sec:Esync}).  For the $E/point$ analysis shown
in the lower panels, the second component models polarized point sources, assuming
$E = B$ spectra rising as $l(l+1)C_l \propto l^2$ and $\beta = -2$
(see \S\ref{sec:Epoint}).
The CMB $E$ component is given in units of amplitude relative to the concordance
model, as before.  For comparison, the amplitudes of the second components are
plotted in units relative to $E=B=8.5 \mu K^2$ at $l=300$, which is the amplitude of
the concordance model $E$ spectrum at that angular scale.
(The $E/sync$ plot shows zero likelihood in the region where unphysical negative amplitudes
for the second component cause the covariance to be non positive-definite).
It is evident that the
level of foreground contamination in either of these models is constrained
to be well below the level of the observed $E$ signal.
}
\label{fig:2_comp}
\end{figure*}

Here and in the next section, we describe two new
analyses using two-component models to constrain the level of
foreground emission in the data set.
In both of these analyses, the first component is the same CMB $E$ power spectrum
as in the $E/B$ concordance-shape polarization analysis, with 
frequency spectral index $\beta = 0$. For the $E/sync$ analysis, the second component is 
defined to be 
a model of polarized synchrotron emission
with $E = B$, flat power spectra (i.e., $l(l+1)C_l \propto \rm{constant}$)
and a temperature spectral index $\beta = -2.9$.  The results of this
analysis, given in Table \ref{tab:cxpolar_like} and shown in Figure~\ref{fig:2_comp},
show that the amplitude of the CMB $E$ component remains consistent with the concordance model,
while the amplitude of the synchrotron component is consistent with zero.
The $95\%$ upper limit on the synchrotron component, $E=B=0.91 \mu K^2$, is well below
the level of the observed $E$ signal, 
implying, as discussed
in \S\ref{sec:discussion}, that there is no significant contamination from
foreground synchrotron emission in the data set.

\subsubsection{$E/point$}

\label{sec:Epoint}

A two-component model was also used to test the extent to which the
signal in our data could be due to point sources.  For this $E/point$
analysis, the second component is a model for polarized point sources,
with $E = B$ and $l(l+1)C_l \propto l^2$ power spectra, for which we
have conservatively assumed $\beta = -2$.  The results are given in
Table \ref{tab:cxpolar_like} and shown in Figure~\ref{fig:2_comp}.
The amplitude of the CMB $E$ component is again consistent with the
concordance model, while the point source model component is consistent
with zero.  The $95\%$ upper limit on the point source component,
$E=B=0.98 \mu K^2$ at $l=300$, is well below the observed signal
level.  We conclude that the presence of point sources consistent with
the data has at most a small effect on the polarization results (see also
\S\ref{sec:discussion}).

\subsubsection{Scalar/Tensor}

To constrain T/S directly from the polarization data, 
we perform a two-component likelihood analysis, using for one
component the 
concordance $E$ shape with $B=0$ for the scalars, and for the
other component the
predicted shapes of $E_T$ and $B_T$ for the tensors.
In principle, because the scalar $B$-mode spectrum is zero, this approach avoids
the fundamental sample variance limitations arising from using the temperature spectrum
alone.  However, the results of the $E5$/$B5$ analysis above
indicate that we can place only upper limits on the $E$ and $B-$mode polarization at
the angular scales most relevant ($l \lesssim 200$) for the tensor spectra.  It is therefore
not surprising that our limits on T/S derived from the polarization spectra,
as reported in Table \ref{tab:cxpolar_like}, remain quite weak.

\begin{table*}
\caption{\label{tab:cxpolar_like}Results of Likelihood Analyses from Polarization Data}
\scriptsize%
\begin{tabular}
[c]{llcrrrrrl}%
		&		&					&		&                                 	& \multicolumn{2}{c}{68\% interval} \\
\rule [-2mm]{0mm}{4mm}
analysis	& parameter 	& $l_{\rm{low}}-l_{\rm{high}}$	& ML est. 	&$\left(F^{-1}\right)_{ii}^{1/2}$ error	& lower	& upper	& U.L.(95\%)	& units\\
\hline\hline
E/B		& E 		& $-$ 		 			&    0.73  	& $\pm0.21$ 				& 0.53	& 0.95	& $-$		& fraction of concordance E\\
		& B 		& $-$ 		 			&    0.03  	& $\pm0.10$ 				& -0.05	& 0.12	& 0.25		& fraction of concordance E\\
\hline																			 
E5/B5 		& E1 		& $28-245$ 	        		&    0.06  	& $\pm0.8$ 				& -0.58	& 1.28	& 3.6 		& $\muK^{2}$\\
		& E2 		& $246-420$ 	 			&   10.5  	& $\pm4.1$ 				&  6.3 	& 16.6	& $-$ 		& $\muK^{2}$\\
		& E3 		& $421-596$ 	 			&    3.6   	& $\pm4.7$ 				& -2.1 	&  8.8	& 15.2		& $\muK^{2}$\\
		& E4 		& $597-772$ 	 			&   22.1  	& $\pm13.0$ 				&  9.8	& 38.8	& $-$ 		& $\muK^{2}$\\
\vspace{1mm}	& E5 		& $773-1050$ 	 			&   42.2  	& $\pm40.0$ 				&  4.5 	& 96.9	& 183.4		& $\muK^{2}$\\
		& B1 		& $28-245$ 	 			&  -0.37 	& $\pm0.55$				& -0.81	& 0.55	& 2.12		& $\muK^{2}$\\
		& B2 		& $246-420$ 	 			&   1.0		& $\pm1.7$ 				& -0.57	& 3.1	& 6.45		& $\muK^{2}$\\
		& B3 		& $421-596$ 	 			&   1.7		& $\pm3.8$ 				& -1.6	&  5.6	& 11.4		& $\muK^{2}$\\
		& B4 		& $597-772$ 	 			&  -1.6		& $\pm6.5$ 				& -6.8	&  5.7	& 16.4		& $\muK^{2}$\\
		& B5 		& $773-1050$ 	 			& -26.7		& $\pm17.9$ 				& -42.9	& -7.0	& 49.3 		& $\muK^{2}$\\
\hline																			 
E/$\beta_E$ 	& E 		& $-$ 		 			& 0.73  	& $\pm0.21$ 				& 0.52	& 0.93	& $-$		& fraction of concordance E\\
		& $\beta_E$ 	& $-$ 		 			& 1.39  	& $\pm1.56$ 				& -0.21	& 2.92	& $-$		& temperature spectral index\\
\hline																			 
E/sync  	& E 		& $-$ 		 			&  0.76  	& $\pm0.21$ 				&  0.56	& 0.99	& $-$		& fraction of concordance E\\
		& sync  	& $-$ 		 			& -0.24 	& $\pm0.28$ 				& -0.49	& 0.21	& 0.91		& $\muK^{2}$\\
\hline																			 
E/point 	& E 		& $-$ 		 			&  0.76  	& $\pm0.23$ 				&  0.55	& 0.99	& $-$		& fraction of concordance E\\
		& point 	& $-$ 		 			& -0.14 	& $\pm0.50$ 				& -0.51	& 0.30	& 0.98		& $\muK^{2}$ at $l=300$\\
\hline																			 
Scalar/Tensor 	& S 		& $-$ 		 			&  0.75   	& $\pm0.22$				&  0.55 & 1.00	& $-$		& fraction of concordance S\\
		& T 		& $-$ 		 			& -5.70		& $\pm7.5$ 				& -12.0	& 4.5 	& 24.9		& T/(S=1)\\
\hline
\end{tabular}

\vspace{12pt}
\begin{tabular}{cccccccccc}
\multicolumn{10}{c}{\rule [-2mm]{0mm}{4mm}parameter correlation matrices}\\

\hline\hline
\\
$\phm{-}{\rm E1}$&$\phm{-}{\rm E2}$&$\phm{-}{\rm E3}$&$\phm{-}{\rm E4}$&$\phm{-}{\rm E5}$&$\phm{-}{\rm B1}$&$\phm{-}{\rm B2}$&$\phm{-}{\rm B3}$&$\phm{-}{\rm B4}$&$\phm{-}{\rm B5}$\\
\hline
\vspace{-6pt}
\\
$\phm{-}1       $ &$-0.137         $ &$\phm{-}0.015   $ &$      -0.002   $ &$\phm{-}0.000   $ &$      -0.241   $ &$\phm{-}0.044   $ &$      -0.004   $ &$\phm{-}0.001   $ &$\phm{-}0.000 $\\ 
                  &$\phm{-}1       $ &$      -0.110   $ &$\phm{-}0.011   $ &$      -0.001   $ &$\phm{-}0.026   $ &$      -0.076   $ &$\phm{-}0.005   $ &$      -0.001   $ &$\phm{-}0.000 $\\ 
                  &                  &$\phm{-}1       $ &$      -0.104   $ &$\phm{-}0.011   $ &$      -0.003   $ &$\phm{-}0.008   $ &$      -0.026   $ &$\phm{-}0.003   $ &$\phm{-}0.000 $\\ 
                  &                  &                  &$\phm{-}1       $ &$      -0.108   $ &$\phm{-}0.000   $ &$      -0.001   $ &$\phm{-}0.002   $ &$      -0.017   $ &$\phm{-}0.002 $\\ 
                  &                  &                  &                  &$\phm{-}1       $ &$\phm{-}0.000   $ &$\phm{-}0.000   $ &$\phm{-}0.000   $ &$\phm{-}0.002   $ &$      -0.016 $\\ 
                  &                  &                  &                  &                  &$\phm{-}1       $ &$      -0.228   $ &$\phm{-}0.024   $ &$      -0.003   $ &$\phm{-}0.000 $\\ 
                  &                  &                  &                  &                  &                  &$\phm{-}1       $ &$      -0.103   $ &$\phm{-}0.014   $ &$      -0.002 $\\ 
                  &                  &                  &                  &                  &                  &                  &$\phm{-}1       $ &$      -0.132   $ &$\phm{-}0.019 $\\ 
                  &                  &                  &                  &                  &                  &                  &                  &$\phm{-}1       $ &$      -0.142 $\\ 
                  &                  &                  &                  &                  &                  &                  &                  &                  &$\phm{-}1     $\\

\hline
\rule [0mm]{0mm}{4mm}
\vspace{-4pt}
$\phm{-}{\rm E}$&$\phm{-}{\rm B}$ & $\phm{-}{\rm E}$&$\phm{-}\beta_E$ & $\phm{-}{\rm E}$&$\phm{-}{\rm sync}$ & $\phm{-}{\rm E}$&$\phm{-}{\rm point}$ & $\phm{-}{\rm S}$&$\phm{-}{\rm T}$ \\
\multicolumn{2}{c}{\hrulefill}&\multicolumn{2}{c}{\hrulefill}&\multicolumn{2}{c}{\hrulefill}&\multicolumn{2}{c}{\hrulefill}&\multicolumn{2}{c}{\hrulefill}\\

$\phm{-}1$ & $-0.048$   & $\phm{-}1$ & $      -0.26$ & $\phm{-}1$ & $      -0.26$ & $\phm{-}1$ & $      -0.43$ & $\phm{-}1$ & $      -0.27$ \\
           & $\phm{-}1$ &            & $\phm{-}1$    &            & $\phm{-}1$    &            & $\phm{-}1$    &            & $\phm{-}1$ \\

\end{tabular}

\normalsize
\end{table*}

\subsection{Temperature Data Analyses and Results for the $T$ Spectrum}

\subsubsection{$T$/$\beta_T$}
\label{sec:tspecind}

In this analysis, we use the bandpower shape of the concordance $T$
spectrum, with the amplitude parameter expressed in units relative to
that spectrum.  The spectral index is relative to the CMB, so that
$\beta = 0$ corresponds to a 2.73~K Planck spectrum.  The amplitude of
$T$ has a maximum likelihood value of 1.23 (1.13 to 1.35), and the
spectral index $\beta_T = 0.11$ ($-0.02$ to 0.24).  While the
uncertainty in the temperature amplitude is dominated by sample
variance, the uncertainty in the spectral index is limited only by the
sensitivity and fractional bandwidth of DASI.  Note that these results
and those for $T5$ below have not been corrected for contributions
from residual point sources (see \papertwo).

\subsubsection{$T5$}
\label{sec:T5}

In the third panel of Figure~\ref{fig:fivebandTEBTE}, we present the
results of an analysis using five, piecewise-continuous, flat
bandpowers to characterize the temperature spectrum.  Although these
results are entirely dominated by the sample variance in the
differenced field, they are consistent with the far more sensitive
temperature power spectrum described in \papertwo.  We include them
here and in \polres, primarily to emphasize that DASI makes
measurements simultaneously in all four Stokes parameters.

\begin{table*}
\caption{\label{tab:copolar_like}Results of Likelihood Analyses from Temperature Data}
\footnotesize%
\begin{tabular}
[c]{llcrrrrl}%
		&		&					&		&                                 	& \multicolumn{2}{c}{68\% interval} \\
\rule [-2mm]{0mm}{4mm}
analysis	& parameter 	& $l_{\rm{low}}-l_{\rm{high}}$	& ML est. 	&$\left(F^{-1}\right)_{ii}^{1/2}$ error	& lower	& upper	& units\\
\hline\hline																	
T/$\beta_T$ 		& T 		& $-$ 		 			& 1.23  	& $\pm0.10$ 				&  1.13	& 1.35	& fraction of concordance T\\
		& $\beta_T$ 		& $-$ 		 			& 0.11   	& $\pm0.10$ 				& -0.02	& 0.24	& temperature spectral index\\
\hline						
T5 		& T1 		& $28-245$ 	        		& 5970  	& $\pm1460$				& 4480     &   7840 & $\muK^{2}$\\
		& T2 		& $246-420$ 	 			& 1840  	& $\pm410$ 				& 1460     &   2300 & $\muK^{2}$\\
		& T3 		& $421-596$ 	 			& 3120  	& $\pm530$ 				& 2560     &   3750 & $\muK^{2}$\\
		& T4 		& $597-772$ 	 			& 2180  	& $\pm460$ 				& 1670     &   2690 & $\muK^{2}$\\
		& T5 		& $773-1050$ 	 			& 3650  	& $\pm1080$ 				& 2660     &   4960 & $\muK^{2}$\\
\hline
\end{tabular}

\vspace{12pt}
\begin{tabular}{cccccp{20pt}cc}
\multicolumn{8}{c}{\rule [-2mm]{0mm}{4mm}parameter correlation matrices}\\
\hline\hline
\\
$\phm{-}{\rm T1}$&$\phm{-}{\rm T2}$&$\phm{-}{\rm T3}$&$\phm{-}{\rm T4}$&$\phm{-}{\rm T5}$ && \phm{-}T & $\phm{-}\beta_T$ \\ 
\cline{1-5} \cline{7-8}
 $\phm{-}1$&  $-0.104$&$\phm{-}0.004$&   $-0.004$&    $-0.002$  && $\phm{-}1$& $0.020$\\ 
           &$\phm{-}1$&      $-0.090$&   $-0.012$&    $-0.011$  &&           &     $1$\\ \cline{7-8} 
           &          &    $\phm{-}1$&   $-0.109$&    $-0.011$ \\ 
           &          &              & $\phm{-}1$&    $-0.142$  \\ 
           &          &              &           &  $\phm{-}1$  \\ 
\cline{1-5}
\end{tabular} 

\normalsize
\end{table*}

\subsection{Joint Analyses and Cross Spectra Results: \te, \tb\ and \eb}
\subsubsection{\tete}
\label{sec:TETE}

\begin{figure*}[t]
\begin{center}
\epsfig{file=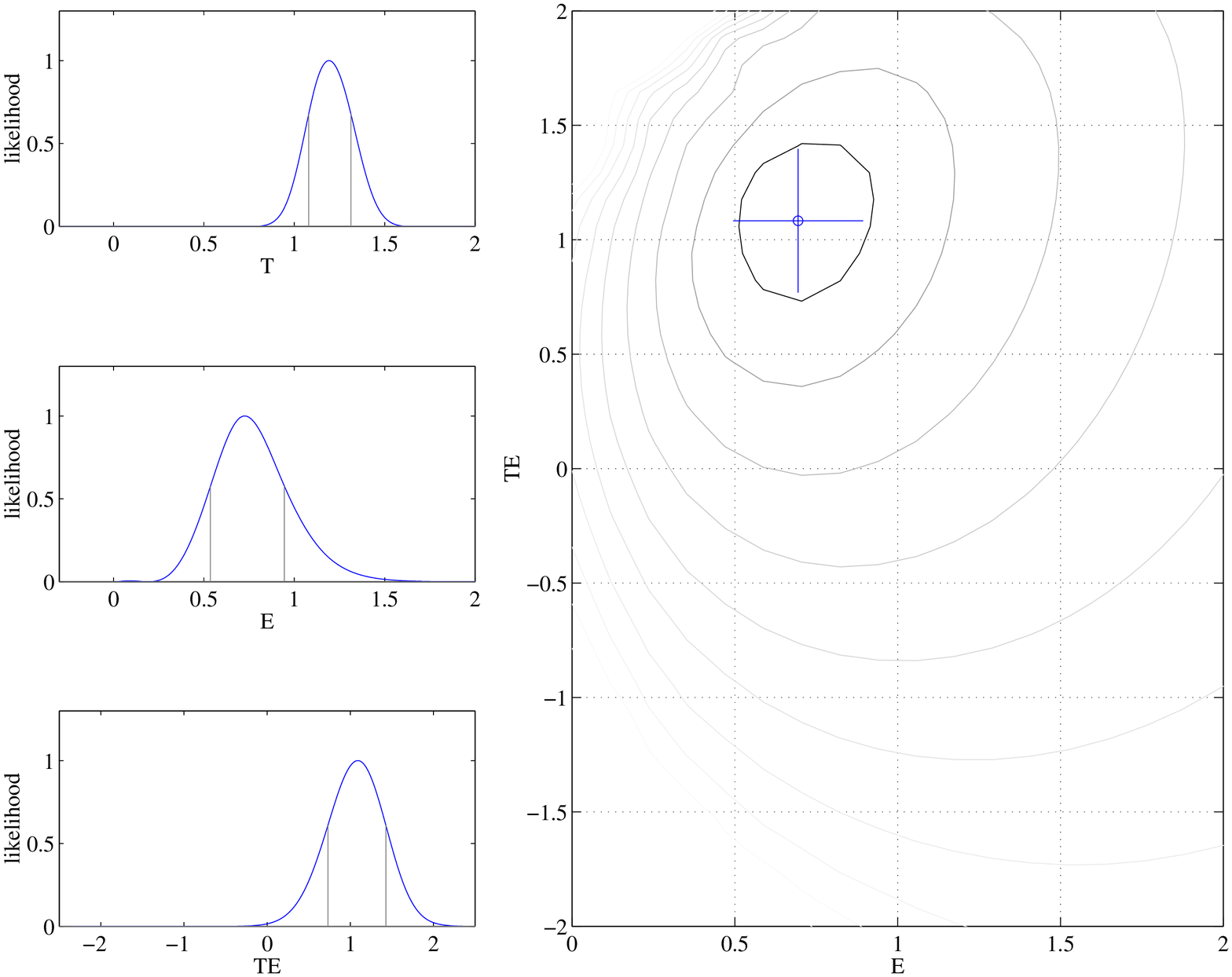,width=\textwidth}
\end{center}
\caption{Results from the 3-parameter shaped bandpower $T$/$E$/\te\ joint analysis.
Spectral shapes are as predicted for the concordance model, 
and the units are relative to that model.
The layout of the plot is analogous to Figure~\ref{fig:EBplot}.
The two dimensional distribution in the right panel is marginalized
over the $T$ dimension.
}
\label{fig:TETEplot}
\end{figure*}

In Figure \ref{fig:TETEplot}, we present results of a three-parameter
single-bandpower analysis of the $T$ and $E$ spectra, and the \te\
cross correlation spectrum, using bandpower shapes from the
concordance model.  The $T$ and $E$ constraints are quite similar to
those from the $E$/$B$, $E$/$\beta_E$ and $T$/$\beta_T$ analyses
described above, as expected.  The additional result here is the \te\
correlation, which has a maximum likelihood value of 1.08 (0.73 to
1.43).  The data are in excellent agreement with concordance model
expectations.

For the no \te-correlation hypothesis, we find a likelihood ratio
$\Lambda({\it TE}=0) = 4.20$ with a PTE of 0.0038; the no correlation
hypothesis is rejected at 2.90$\sigma$, a considerable improvement
over the \te\ results presented in \polres.  Note that under the ${\it
TE}=0$ hypothesis, negative and positive \te\ are equally likely, so
the probability of falsely detecting a positive \te\ correlation at
this level is PTE = 0.0019, or 3.11$\sigma$.

\subsubsection{\tetefive}
\label{sec:TETE5}

This is a seven-parameter analysis using single shaped band
powers for $T$ and $E$, and five piecewise-continuous, flat bandpowers for the \te\ cross correlation. The $B$-mode polarization is explicitly set to zero
for this analysis. The $T$ and $E$ constraints are again similar to the values for the other
analyses where these parameters appear. The interesting 
result here is the \te\ spectrum,  
shown in the bottom panel of
Figure~\ref{fig:fivebandTEBTE}. The bandpower results show 
considerable improvement
from the results presented in \polres, and trace
the specific predictions of the 
concordance model.

\subsubsection{\tebtetbeb}

Our last analysis is a six shaped-bandpower analysis for the three
individual spectra $T$, $E$ and $B$, together with the three possible
cross-correlation spectra \te, \tb\ and \eb.  Although there is no
evidence for any $B$-mode signal, we include the $B$ cross-spectra for
completeness.  Since the \tb\ and \eb\ spectra are expected from
theory to be zero, we preserve the symmetry of the analysis between
$E$ and $B$ by parameterizing them in terms of the \te\ and $E$
shapes, respectively.  The results for $T$, $E$, $B$ and \te\ are again
similar to those obtained in the other analyses. As expected under the
concordance model, there is no detection of \eb\ or \tb.

\begin{table*}
\caption{\label{tab:joint_like}Results of Likelihood Analyses From Joint Temperature-Polarization Data}
\scriptsize%
\begin{tabular}
[c]{llcrrrrl}%
		&		&					&		&                                 	& \multicolumn{2}{c}{68\% interval} \\
\rule [-2mm]{0mm}{4mm}
analysis	& parameter 	& $l_{\rm{low}}-l_{\rm{high}}$	& ML est. 	&$\left(F^{-1}\right)_{ii}^{1/2}$ error	& lower	& upper	& units\\
\hline\hline																	
T/E/TE		& T 		& $-$ 		 			&   1.17  	& $\pm  0.10$ 				& 1.08 & 1.31 & fraction of concordance T\\
		& E 		& $-$ 		 			&   0.69  	& $\pm  0.20$ 				& 0.54 & 0.94 & fraction of concordance E\\
		& TE		& $-$ 		 			&   1.08  	& $\pm  0.31$ 				& 0.74 & 1.41 & fraction of concordance TE\\
\hline																		
T/E/TE5		& T 		& $-$ 		 			&   1.17  	& $\pm  0.10$ 				&   1.10 &   1.32 & fraction of concordance T\\
\vspace{1mm}	& E  		& $-$ 		 			&   0.78  	& $\pm  0.21$ 				&   0.64 &   1.09 & fraction of concordance E\\
      		& TE1 		& $28-245$ 	        		& -45.9 	& $\pm 24.3$ 				& -69.3  & -11.2  & $\muK^{2}$\\
		& TE2 		& $246-420$ 	 			& 121.9  	& $\pm 30.7$ 				&  82.9  & 159.8  & $\muK^{2}$\\
		& TE3 		& $421-596$ 	 			& -19.0 	& $\pm 41.0$ 				& -62.2  &  23.0  & $\muK^{2}$\\
		& TE4 		& $597-772$ 	 			& -91.8 	& $\pm 60.5$ 				& -152.5 & -22.9  & $\muK^{2}$\\
            	& TE5 		& $773-1050$ 	 			& -14.4  	& $\pm129.3$ 				& -117.8 &  83.4  & $\muK^{2}$\\
\hline																		
T/E/B/TE/TB/EB	& T 		& $-$ 		 			&  1.17		& $\pm  0.10$ 				& 1.08	& 1.30	& fraction of concordance T\\
              	& E 		& $-$ 		 			&  0.68		& $\pm  0.20$ 				& 0.56	& 0.98	& fraction of concordance E\\
              	& B 		& $-$ 		 			&  0.02		& $\pm  0.09$ 				&-0.03	& 0.15	& fraction of concordance E\\
              	& TE		& $-$ 		 			&  1.17		& $\pm  0.30$ 				& 0.84	& 1.52	& fraction of concordance TE\\
              	& TB		& $-$ 		 			&  0.14		& $\pm  0.24$ 				&-0.16	& 0.35	& fraction of concordance TE\\
              	& EB		& $-$ 		 			& -0.19		& $\pm  0.10$ 				&-0.31	&-0.10	& fraction of concordance E\\
\hline
\end{tabular}

\vspace{12pt}
\begin{tabular}{cccccccp{20pt}cccccc}
\multicolumn{14}{c}{\rule [-2mm]{0mm}{4mm}parameter correlation matrices}\\
\hline\hline
\\
$\phm{-}{\rm T}$&$\phm{-}{\rm E}$&$\phm{-}{\rm TE1}$&$\phm{-}{\rm TE2}$&$\phm{-}{\rm TE3}$&$\phm{-}{\rm TE4}$&$\phm{-}{\rm TE5}$&&$\phm{-}{\rm T}$&$\phm{-}{\rm E}$&$\phm{-}{\rm B}$&$\phm{-}{\rm TE}$&$\phm{-}{\rm TB}$&$\phm{-}{\rm EB}$\\ 
\cline{1-7} \cline{9-14}
$\phm{-}1$&$\phm{-}0.068$&  $-0.159$&$\phm{-}0.303$&      $-0.037$&      $-0.126$&$-0.010$ && $\phm{-}1$&$\phm{-}0.046$&      $-0.014$&$\phm{-}0.299$&$\phm{-}0.044$&$\phm{-}0.018$\\ 
          &    $\phm{-}1$&  $-0.146$&$\phm{-}0.484$&      $-0.045$&      $-0.150$&$\phm{-}0.001$ &&           &    $\phm{-}1$&$\phm{-}0.009$&$\phm{-}0.407$&      $-0.079$&      $-0.310$\\ 
          &              &$\phm{-}1$&      $-0.127$&$\phm{-}0.013$&$\phm{-}0.039$&$\phm{-}0.001$ &&           &              &$\phm{-}1$    &$      -0.018$&$\phm{-}0.082$&      $-0.345$\\ 
          &              &          &    $\phm{-}1$&      $-0.083$&      $-0.099$&      $-0.002$ &&           &              &              &    $\phm{-}1$&      $-0.216$&      $-0.014$\\ 
          &              &          &              &    $\phm{-}1$&      $-0.045$&$\phm{-}0.004$ &&           &              &              &              & $\phm{-}1$   &$\phm{-}0.295$\\ 
          &              &          &              &              &    $\phm{-}1$&      $-0.058$ &&           &              &              &              &              &$\phm{-}1$    \\ \cline{9-14} 
          &              &          &              &              &              &    $\phm{-}1$  \\ 
\cline{1-7}
\\
$\phm{-}{\rm T}$&$\phm{-}{\rm E}$&$\phm{-}{\rm TE}$  \\ 
\cline{1-3}
 $\phm{-}1$&$\phm{-}0.036$&$\phm{-}0.276$\\ 
           &    $\phm{-}1$&$\phm{-}0.372$\\ 
           &              &    $\phm{-}1$\\ 
\cline{1-3}
\end{tabular} 

\vspace{12pt}
\begin{tabular}{ccc}
\end{tabular} 

\normalsize
\end{table*}

\section{Discussion}
\label{sec:discussion}

The analysis of the \threeyear\ DASI data set supports all of the
results presented in \polres, in particular the detection of \emode\
CMB polarization consistent with predictions of the standard
cosmological model, and inconsistent with any significant
contamination from foreground emission. The increased sensitivity of
the combined data set permits a more precise characterization of the
detected polarization signal, and increasingly sensitive tests for the
presence of foreground contamination.

As discussed in \S\ref{sec:consistency}, simple consistency tests with
no assumptions about the underlying model show a high confidence
detection of polarized signal.  When the polarized anisotropy is
modeled by a shaped spectrum over the $l$-range spanned by DASI, the data
strongly prefer the concordance model shape for the \emode\ spectrum
(see \S\ref{sec:EB}).  Assuming the concordance model shape, we detect
\emode\ polarization (at $6.3\sigma$ confidence), as well as the \te\
cross-correlation ($2.9\sigma$), both with significantly increased
statistical power over the results in \polres, and at levels consistent
with theoretical expectations. As expected for CMB anisotropies, the
level of the \bmode\ polarization is consistent with zero, with an
upper limit significantly below the \emode\ detection.

The five piecewise-continuous bandpower analysis presented in
\S\ref{sec:E5B5} further demonstrates that the detailed shape of the
\emode\ anisotropy is consistent with the concordance predictions. The
\emode\ bandpowers place a stringent upper limit in the lowest
$l$-range band, with maximum likelihood values in the remaining four bands
that are consistent with the damped concordance model, but
inconsistent with a simple $l(l+1)C_l \propto l^2$ power law.  As shown
in Figure~\ref{fig:fivebandTEBTE} (\S\ref{sec:TETE5}), the \te\
angular power spectrum provides a highly specific test of the spectral
shape, as the predicted \te\ correlation crosses zero five times over
the angular range covered by the DASI measurements.

Of the known foregrounds at 30 GHz, diffuse Galactic synchrotron
emission and radio point sources are of the greatest concern. (As
discussed in \polres, neither free-free nor dust emission, of either
the thermal or spinning variety, is expected to contribute any
significant polarized signal to the DASI data.)  It was for this
reason that the two DASI fields were carefully selected from regions
of exceptionally low emission in the Haslam maps \citep{haslam81},
confirmed at higher frequency by the \wmap\ synchrotron maps to be
clean regions of sky \citep{bennett03b} (see
Figure~\ref{fig:wmap}). The two fields were also selected from those
with no detectable point sources from the original 32 DASI temperature
anisotropy fields.

While no published maps exist of the polarized synchrotron emission
toward our fields, the DASI data themselves set stringent constraints
on the allowed level of diffuse synchrotron. As discussed in
\S\ref{sec:Esync}, a two-component model, designed specifically to
probe for synchrotron emission, finds that the amplitude of a
synchrotron component with $\beta = -2.9$ is consistent with zero,
with an upper limit well below the level of the CMB \emode\ signal.
The \emode\ spectral index constraint, improved from \polres,
indicates that the polarized signal is consistent with CMB; a signal
with the expected synchrotron spectral index $\beta = -2.9$ is
rejected at the 2.7$\sigma$ level.  The detection of the \te\
correlation is strong evidence that the polarized signal shares a
common origin with the total intensity signal, a signal which has been
robustly demonstrated in \papertwo, \polres\ and here to be consistent
with CMB in both its frequency spectral index ($\beta = 0.11$ ($-0.02$
to $0.24$) from \S\ref{sec:tspecind} above) and angular power
spectrum.  Moreover, the pronounced asymmetry of the \emode\ and
\bmode\ levels allowed by DASI is in sharp contrast to the symmetric
predictions for sources of foreground emission \citep{zaldarriaga01}.

Since no catalog of faint polarized sources exists over our observing
region, and it is not possible to identify from total intensity which
point sources will have the strongest polarized flux, it is
impracticable to project sources out of the polarization analyses.
Nevertheless we can estimate the level of residual contamination in
the data.  A new, two-component analysis performed on the \threeyear\
data set finds that the amplitude of a point-source component, with a
conservative spectral index $\beta = -2$, is consistent with zero,
with an upper limit well below the level of the CMB \emode\ signal
(see \S\ref{sec:Epoint}).

We analyzed maps of the differenced fields directly for evidence of
individual sources appearing above the noise floor.  The pixel value
distributions are consistent with instrument noise plus a Gaussian
signal, and show no sign of excess tails due to discrete sources.  The
sensitivity of our maps is such that point sources with
beam-attenuated polarized flux greater than 15 mJy would have been
clearly detected.

In \polres, we also presented a Monte Carlo analysis of the expected
point-source contribution, based on available knowledge of the source
populations.  The number-flux counts used in the simulation
(determined from the 2000 DASI temperature data (see \polres)) are
consistent with the \wmap\ results on $dN/dS$ \citep{bennett03b}, and
while there is a dearth of data concerning the distribution of source
polarization fractions at 30~GHz, a recent 18.5~GHz study
\citep{ricci04} reports a distribution similar to the one assumed in
our calculations.  These simulations suggest a mean bias of the $E$
parameter of 0.04 with a standard deviation of 0.05.  We therefore
conclude that there is neither the expectation of, nor evidence in the
data for, any significant contamination by polarized point sources.

\section{Conclusion}
\label{sec:conclusion}

DASI has detected \emode\ CMB polarization with high confidence ($6.3
\sigma$ with the addition of the new data), and has
detected the \te\ correlation at the $2.9\sigma$ level.  The shape and
amplitude of the detected \emode\ angular power spectrum are in
agreement with theoretical predictions based on the CMB temperature
measurements.  These results lend strong support to the underlying
theoretical framework for the generation of CMB anisotropy and point
toward a promising future for the field --- a future which includes
the detailed characterization of the \emode\ spectrum, the detection
on small scales of \bmode s from gravitational lensing, and the
tantalizing prospect of constraining inflationary models via the
detection of large-scale \bmode s from primordial gravitational waves.

The DASI polarization results demonstrate the beauty of using
interferometric techniques to provide well-matched filters for $E$-
and \bmode\ measurements, and the inherent power of interferometers to
reject uncorrelated noise and control systematics. The DASI results
would continue to improve with increased integration time, however the
sensitivity required for the detection of the \bmode\ signatures
demands large arrays, beyond the current scaling limitations of
existing correlators.  While there are currently no plans to upgrade
DASI as an interferometer, advances in technology may once again make
interferometry an attractive choice for pursuing the extraordinary
control of systematics required by the next generation of CMB
polarization experiments.

\bigskip 

We thank Ben Reddall, Eric Sandberg and Allan Day
for their dedicated and professional support to the DASI
polarization experiment while they wintered at the 
NSF Amundsen-Scott South Pole research station.
We are indebted to Bill Holzapfel and Mark Dragovan, whose
early contributions to the DASI experiment were essential
to the success of the program, and to the CBI team led
by Tony Readhead, in particular to John Cartwright, Steve Padin
and Martin Shepherd.
We are grateful for expert technical assistance from 
Jacob Kooi,
Charlie Kaminski,
Ellen LaRue,
Bob Pernic,
Bob Spotz and
Mike Whitehead. We thank the U.S.\ Antarctic Program and
the Raytheon Polar Services Corporation for their support
of the project.
The DASI project is supported by NSF grant OPP-0094541. This research
was also supported in part by NSF grant PHY-0114422 to the Kavli
Institute for Cosmological Physics, a NSF Physics Frontier Center.
NWH acknowledges support from a Charles H. Townes Fellowship
at the U.C.B. Space Sciences Laboratory. JMK acknowledges
support from a Millikan Fellowship at Caltech.

\bibliographystyle{apj}
\bibliography{jc}
\end{document}